\documentclass[aps,prd,groupedaddress]{revtex4-1}

\usepackage{graphicx}
\usepackage{dcolumn}
\usepackage{bm}
\usepackage{amsmath}
 
\def\mbi#1{\mbox{\bfseries\itshape #1}} 

\newcommand{\advast}{\rm Adv.~Astron.~}

\newcommand{\aap}{\rm Astron.~\&~Astrophys.~}
\newcommand{\aapr}{\rm Astron.~Astrophys.~Rev.~}

\newcommand{\apjl}{\rm Astrophys.~J.~Lett.~}

\newcommand{\jcap}{\rm J. Cosmol. Astropart. Phys.~}

\newcommand{\mnras}{\rm Mon.~Not.~R.~Astron.~Soc.~}

\newcommand{\plb}{\rm Phys.~Lett.~B~}
\newcommand{\physrep}{\rm Phys.~Rep.~}

\newcommand{\D}{\displaystyle}

\begin{document}


\title{Impact of a primordial magnetic field on cosmic microwave background $B$ modes with weak lensing}

\author{Dai G. Yamazaki$^{1,2}$}
 \email{yamazaki.dai@nao.ac.jp}
\affiliation{%
$^{1}$Ibaraki University, 2-1-1, Bunkyo, Mito, 310-8512, Japan}%
\affiliation{%
$^{2}$National Astronomical Observatory of Japan, Mitaka, Tokyo 181-8588, Japan}%
\date{\today}

\begin{abstract}
We discuss the manner in which the primordial magnetic field (PMF) suppresses the cosmic microwave background (CMB) $B$ mode due to the weak-lensing (WL) effect. The WL effect depends on the lensing potential (LP) caused by matter perturbations, the distribution of which at cosmological scales is given by the matter power spectrum (MPS). Therefore, the WL effect on the CMB $B$ mode is affected by the MPS. Considering the effect of the ensemble average energy density of the PMF, which we call ``the background PMF,'' on the MPS, the amplitude of MPS is suppressed in the wave number range of
$k>0.01~h$ Mpc$^{-1}$.
The MPS affects the LP and the WL effect in the CMB $B$ mode; however, the PMF can damp this effect. Previous studies of the CMB $B$ mode with the PMF have only considered the vector and tensor modes. These modes boost the CMB $B$ mode in the multipole range of 
$\ell > 1000$, whereas the background PMF damps the CMB $B$ mode owing to the WL effect in the entire multipole range. The matter density in the Universe controls the WL effect. Therefore, when we constrain the PMF and the matter density parameters from cosmological observational data sets, including the CMB $B$ mode, we expect degeneracy between these parameters. The CMB $B$ mode also provides important information on the background gravitational waves, inflation theory, matter density fluctuations, and the structure formations at the cosmological scale through the cosmological parameter search. If we study these topics and correctly constrain the cosmological parameters from cosmological observations including the CMB $B$ mode, we need to correctly consider the background PMF.
\end{abstract}
\maketitle 
\section{Introduction \label{sec:introduction}}
The polarization isotropy of the cosmic microwave background (CMB) has the odd parity (curl) component. This is called the ``B'' mode. The gravitational wave background (GWB) and weak lensing (WL) are considered the sources of the CMB $B$ mode. The GWB contains information for the inflation, and WL depends on the cosmological matter fluctuations. Therefore, investigating the CMB $B$ mode is expected to yield important information on inflation theory, cosmological matter density fluctuations, and the structure formation at cosmological scales.

The WL effect dominates the CMB $B$ mode on hundreds of multipoles. In this range, several observational projects (ACTpol \cite{2017JCAP...06..031L}, SPTpol \cite{2015ApJ...807..151K}, PORABEAR \cite{2017ApJ...848..121T}, and BICEP2/Keck \cite{2016PhRvL.116c1302B})  have provided interesting data sets for the CMB $B$ mode. The next generation plans for these projects are expected to advance the work on the CMB $B$ mode, including the WL effect.

Magnetic fields with strength on the order of 1 $\mu$G ($=10^{-6}$ G) are observed in the typical subgalaxy to cluster scale by Faraday rotation and synchrotron emission \cite{2004IJMPD..13.1549G,Wolfe:1992ab,Clarke:2000bz,Xu:2005rb,2010Sci...328...73N,2012ApJ...747L..14V,2012A&ARv..20...54F}.
The primary origin of these magnetic fields is the primordial magnetic field (PMF). PMF is assumed to be homogeneous and stochastic with comoving strength on the order of $1a^{-2}$ nG (1 nG$=10^{-9}$ G), where $a$ is the scale factor.
If this PMF is generated before the recombination and it evolves into the observed magnetic fields by the isotropic collapse of the density fields in the early Universe, then the observed magnetic fields at cosmological scales can be explained  \cite{Grasso:2000wj,2010AdAst2010E..80Y,2011PhR...505....1K,2012PhR...517..141Y}.

There are two kinds of PMF effects on the fluctuations of the CMB and the matter power spectrum (MPS), which relate the spatial distributions of the fluctuations of the matter densities with the wave number. The first effect is due to the first perturbation source from the PMF in the linear perturbative equations. Therefore, this PMF effect is called ``the perturbative PMF effect'' in this paper. The perturbative PMF effect generates fluctuations of CMB and MPS in smaller scales corresponding to $k \ge 0.1~h$Mpc$^{-1}$ \cite{2006PhRvD..74l3518Y,2008PhRvD..77d3005Y,2008PhRvD..78b3510F, 2010PhRvD..81b3008Y},where $h$ is the Hubble parameter in units of 100 km/s/Mpc.
The PMF effects have been extensively studied \cite{Durrer:1999bk,Grasso:2000wj,Mack:2001gc,Kosowsky:2004zh,Kahniashvili:2005xe,Dolgov:2005ti,Giovannini:2007qn,2008PhRvD..77d3005Y,Paoletti:2008ck,Sethi:2008eq,Kojima:2008rf,2008PhRvD..78f3012K,Giovannini:2008aa,Shaw:2009nf,2010AdAst2010E..80Y,2011PhR...505....1K,2011PhRvD..84l3006Y,2012PhR...517..141Y} and constrained by using the cosmological observation data sets \cite{Yamazaki:2004vq,Lewis:2004ef,Yamazaki:2006bq,Yamazaki:2008bb,Kahniashvili:2008hx,2010PhRvD..81b3008Y,2010PhRvD..82h3005K,2011PhRvD..83l3533P,2012PhRvD..86d3510S,2013PhRvD..88j3011Y,2013PhLB..726...45P,2013JCAP...11..006S,2016A&A...594A..19P,2017PhRvD..95f3506Z}.
The second effect is due to the ensemble energy density of the PMF because it is considered a nonperturbative source in linear perturbative theory. Therefore, we call it ``the background PMF effect.'' The background PMF changes the features of the CMB \cite{2014PhRvD..89j3528Y} and MPS \cite{2016PhRvD..93d3004Y}.
The peak positions of the CMB temperature fluctuations and the MPS shift to a larger scale because of the background PMF \cite{2014PhRvD..89j3528Y,2016PhRvD..93d3004Y}.
The amplitude of the MPS is also suppressed at smaller scales than the peak position $k_p$, which corresponds to $0.01~h\mathrm{Mpc}^{-1}~<~k~<~0.02~h$Mpc$^{-1}$, by the background PMF \cite{2016PhRvD..93d3004Y}. As mentioned above, the effects of perturbative PMF on the cosmology and the constraints on perturbative PMFs have been studied; however, the effects of background PMF on the CMB $B$ mode and constraints on the background PMF from the CMB $B$ mode have not been studied.

In this paper, we analyze the background PMF effects on the CMB $B$ mode separately from the WL effects for the first time. We also investigate the overall effects of the perturbative and background PMFs on the CMB $B$ mode and discuss the degeneracies between the PMF and the matter density parameters.
We adopt the cosmological parameters determined by the Planck 2015 (TT + lowP + lensing in Table 4 of Ref. \cite{2016A&A...594A..13P}) as follows:
$(\Omega_\mathrm{b}h^2, ~\Omega_\mathrm{CDM}h^2,~n_s,~ \ln (10^{10} A_s), ~H_0, ~\tau) = (0.02226, ~0.1186,~0.9677,~3.062,~67.81,~0.066)$,
 where
 $\Omega_\mathrm{b}$ and $\Omega_\mathrm{CDM} $ are the baryon and the cold dark matter (CDM) density parameters, respectively;
 $n_s$ is the scalar spectral index;
 $A_s$ is the scalar amplitude of the initial fluctuation;
 $H_0$ is the Hubble constant; and 
 $\tau$ is the optical depth.
We modify CAMB code \cite{camb} for computing the lensing potential and the CMB $B$ mode with PMF effects.
\section{Model of PMF}
We use the flat and $\Lambda$ CDM model. 
We assume that the PMF is generated well before the recombination epoch, e.g., the inflation epoch, and that it is stochastic homogeneous, isotropic, and random. The PMF is also assumed ``frozen in'' the cosmological ionized fluid \cite{Mack:2001gc}.
In this case, the comoving strength of the PMF $B(x)$ is conserved at greater than the cutoff scale length $k_\mathrm{max}$, where $k_\mathrm{max}$ is defined by the PMF damping \cite{Jedamzik:1996wp, Subramanian:1997gi,Mack:2001gc}. Therefore the physical strength of the PMF is given by $B(x, a) = B(x)/a^2$, where $a$ is the scale factor.

We adopt a power-law(PL) PMF model \cite{Mack:2001gc,2014PhRvD..89j3528Y,2016PhRvD..93d3004Y}(see Appendix \ref{sec:AI}) and 
the numerical formulation of the PL PMF spectra 
from Refs. \cite{2008PhRvD..77d3005Y,2010AdAst2010E..80Y,2011PhRvD..84l3006Y}.

In the comoving coordinates, the ensemble energy density of the PL PMF generated well before the recombination epoch is defined by Refs. \cite{Subramanian:1997gi,Subramanian:1998fn,2012PhRvD..86l3006Y,2014PhRvD..89j3528Y,2016PhRvD..93d3004Y}(see Appendix \ref{sec:AI}),
\begin{eqnarray}
\rho_\mathrm{MF}
\sim
\frac{1}{8\pi a^4}
\frac{
B^2_\lambda
}
{
   \Gamma
   \left(
      \frac{n_\mathrm{B}+5}{2}
   \right)
}
(\lambda k_\mathrm{max})^{n_\mathrm{B}+3},
\label{eq:BG_PL_PMF_energy_densityII}
\end{eqnarray}
where
$n_\mathrm{B}$ is the spectrum index of the PMF, 
$B_\lambda=|\bm{B}_\lambda|$ is the comoving field strength by smoothing over a Gaussian sphere of radius $\lambda=1$ Mpc ($k_\lambda = 2\pi/\lambda$),
$\Gamma(x)$ is the gamma function, and
$k_\mathrm{max}$ is the cutoff scale and is defined by the PMF damping \cite{Jedamzik:1996wp, Subramanian:1997gi,Mack:2001gc}.
Since the PMF energy density is proportional to the square of the PMF strength, the time evolution of the ensemble energy density is inversely proportional to $a^{4}$ as the energy density of the radiation.

Based on Refs. \cite{2014PhRvD..89j3528Y,2016PhRvD..93d3004Y}, the effective sound speed with the background PMF in the baryon-photon fluid is 
\begin{eqnarray}
c^2_\mathrm{sA}
=
c^2_\mathrm{s}
+
\frac{1}{2}
c^2_\mathrm{A}.
\label{result_c_sA}
\end{eqnarray}
The first term $c_\mathrm{s}$ in Eq. (\ref{result_c_sA}) is the sound velocity of a fluid without PMF and is $c^2_\mathrm{s} = c^2_\mathrm{b\gamma} = 1/\{ 3(1+\frac{3\rho_\mathrm{b}}{4\rho_\gamma}) \} = 1/\{ 3\left(1+R\right) \} $, 
where $\rho_\mathrm{b}$ and $\rho_\gamma$ are the baryon density and photon energy density, respectively, and $R$ is $\frac{3}{4}\frac{\rho_\mathrm{b}}{\rho_\gamma}$. 
The second term $c_\mathrm{A}$ in Eq. (\ref{result_c_sA}) is the Alfven velocity and is given by $c^2_\mathrm{A} = \frac{2\rho_\mathrm{MF} }{\rho_\gamma + \rho_\mathrm{b}}$ \cite{2014PhRvD..89j3528Y,2016PhRvD..93d3004Y}. 

We do not consider any correlations between the PMF and the primary; e.g., the correlation term is ignored, as in Eq. (22) of \cite{2014PhRvD..89j3528Y}.
We use the initial condition model from Refs. \cite{2008PhRvD..77d3005Y,Shaw:2009nf}.
The sound velocity is not effective for determining the initial condition at the sub- and superhorizon. In this study, the expression for the initial condition with the PMF corresponds to that of Refs. \cite{2008PhRvD..77d3005Y,Shaw:2009nf};
we change the elements of the total energy density, e.g., $\rho = \rho_\mathrm{R} + \rho_\mathrm{M}$ to $\rho = \rho_\mathrm{R} + \rho_\mathrm{M} + \rho_\mathrm{MF}$ as in Refs. \cite{2014PhRvD..89j3528Y,2016PhRvD..93d3004Y},
where $\rho_\mathrm{R}$ is the total radiation energy density and $\rho_\mathrm{M}$ is the total matter density in the Universe.

Reference \cite{2016A&A...594A..19P} constrains the upper bound of the magnetic spectral index as $n_\mathrm{B} < -0.31$ (Table 1 in \cite{2016A&A...594A..19P}). In general, physical futures tend to appear near the upper bounds, and analyzing them is, also, useful to discuss constraints on the parameters of the PMF. Therefore, we choose the magnetic spectral indexes which is close to the upper bound.

The wave number $k_\mathrm{MJ}$ derived by the magnetic Jeans scale \cite{Peebles:1980booka,Kim:1994zh,Tashiro:2005ua} with the order of 1 nG is the order of 10 Mpc$^{-1}$, where we should consider nonlinear effects. However, because we study the effects of the PMF on $k < 0.3 h\mathrm{Mpc}^{-1}$, which is much less than $k_\mathrm{MJ}$, we do not consider peculiar effects of the PMF, e.g., nonlinear effects, around and greater than the wave numbers.
\section{The effects of the PMF on the MPS\label{sec:III}}
The paths of the CMB photons from the last scattering surface to our detectors are deflected by the gravitational potentials owing to the inhomogeneous mass distribution in the Universe. This is called the ``weak lensing.'' The WL effect depends on the matter density field. To study the WL effect on the CMB $B$ mode, we should understand the time evolutions of the matter density field at the cosmological scale.
In this section, first, we show the two important effects on them: the Meszaros and the sound-wave effects.
\subsection{Meszaros effect\label{sec:III_1}}
In the radiation-dominated era, the expansion rate of the Universe is proportional to $\sqrt{G\rho_\mathrm{R}}$, while the evolution rate of the fluctuations of the matter density field is proportional to $\sqrt{G\rho_\mathrm{M}}$. 
Here, $G$ is Newton's constant; $\rho_\mathrm{R}$ is the total radiation (the photon, neutrino, and PMF) energy density; and $\rho_\mathrm{M}$ is the total matter (baryon and dark matter) density. 
Since $\sqrt{G\rho_\mathrm{M}}$ is less than $\sqrt{G\rho_\mathrm{R}}$ due to $\rho_\mathrm{R} > \rho_\mathrm{M}$ in the radiation-dominated era, the matter fluctuations in the superhorizon cannot grow.
This is called the ``Meszaros effect'' \cite{1974A&A....37..225M}. 
After the matter-radiation equality time, 
which is determined by the ratio of the total matter density ($\rho_\mathrm{M}$) to the total radiation energy density ($\rho_\mathrm{R}$), the potential field can affect the matter density fields, and the fluctuations of the matter density fields start to evolve in the horizon. 
The matter fluctuations at smaller scales enter the horizon earlier, and they are affected by the Meszaros effect for a longer time. Therefore, the amplitudes of the MPS on the longer wave numbers are more strongly damped by the Meszaros effect.
\subsection{The sound-wave effect\label{sec:III_2}}
From Ref. \cite{Peebles:1980booka}
the evolution of baryon density fluctuation $\delta$ at the horizon scale for the flat and $\Lambda$ CDM model is
\begin{eqnarray}
\delta \propto
\frac{1}{\sqrt{c_\mathrm{S} a}}\exp \left(-ik \int \frac{c_\mathrm{S}}{a} dt\right),
\label{eq_evolution_density}
\end{eqnarray}
where $c_\mathrm{S}$ is the sound speed. 
From this equation, the baryon density fluctuation in the horizon is influenced by the suppression of the sound wave.
The baryon fluctuation affected by sound waves also suppresses the potential fluctuation.
Although the dark matter fluctuation is not directly affected by the sound speed, the potential fluctuation suppressed delays the evolution of the dark matter fluctuation.
The suppressions of the matter density and the potential fluctuations owing to sound waves at smaller scales have been confirmed numerically \cite{Ma:1995ey,1996ApJ...471..542H,Yamamoto:1997qc}.
As with the case of the Meszaros effect, since the matter density fluctuations at smaller scales enter the horizon earlier and the duration of the damping effect on them is longer, they are more suppressed at the smaller scales.
\subsection{The peak position and the damping of the MPS\label{sec:III_3}}
\begin{figure}
\includegraphics[width=1.0\textwidth]{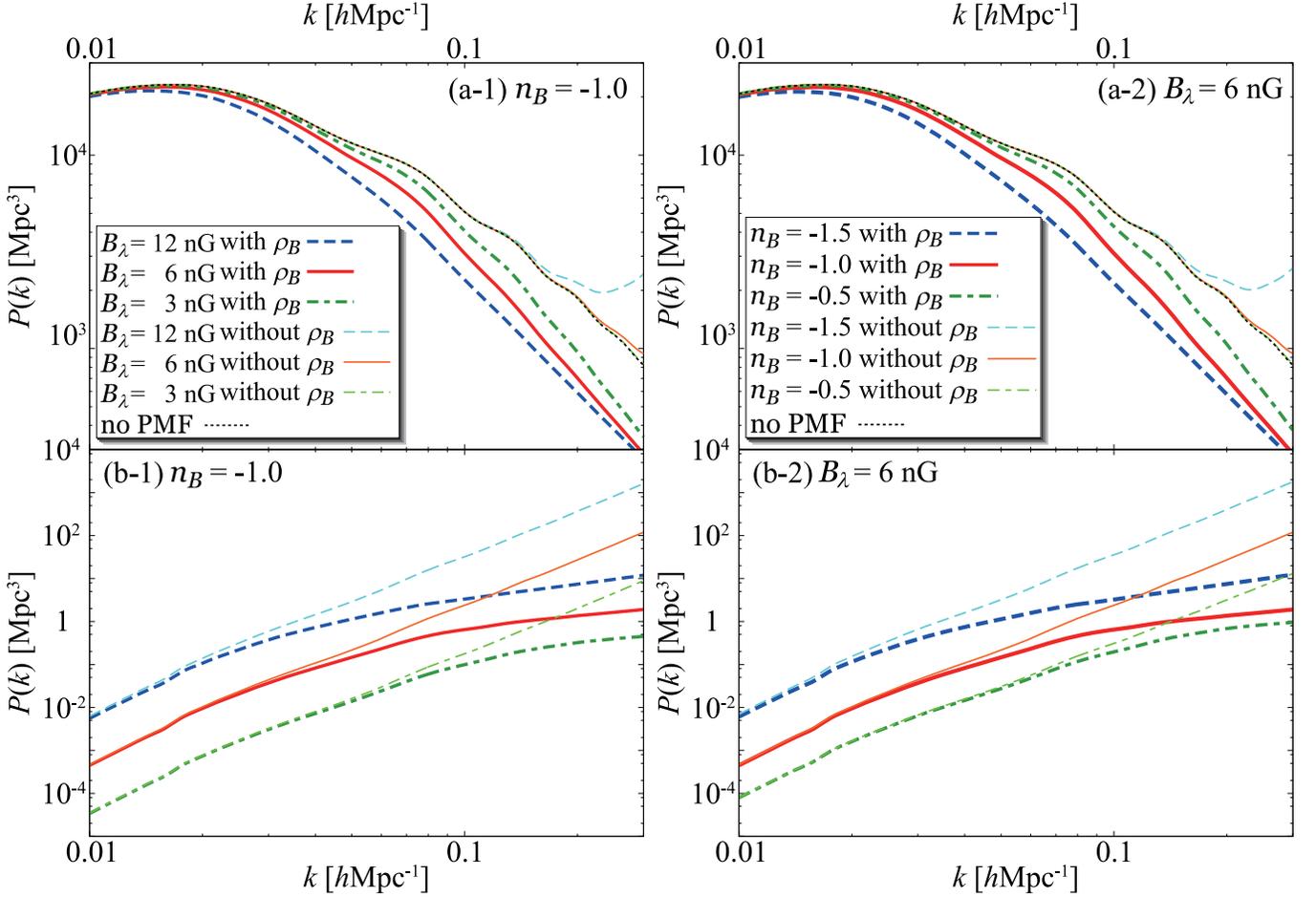}
\caption{\label{fig_supp_1}
PMF effects on the MPS for fixed PMF parameters.
The thin dotted curve denotes the theoretical result without PMF effects and based on the cosmological parameters determined by Planck 2015 (TT + lowP + lensing in Table 4 of Ref. \cite{2016A&A...594A..13P}). 
The power spectral index of the PMF in panels (a-1) and (b-1) is fixed as $n_B = -1.0$, and the other corves in panels (a-1) and (b-1) are the theoretical result with the PMF effects, as shown by the legends in panel (a-1). 
The field strength of the PMF in panels (a-2) and (b-2) is fixed as $B_\lambda = 6.0$ nG, and the other corves in panels (a-2) and (b-2) are the theoretical result with the PMF effects as shown by the legends in panel (a-2).
} 
\end{figure}

Figure \ref{fig_supp_1} shows the MPS. The vertical axis in Fig. \ref{fig_supp_1} indicates the amplitude of the matter density fluctuations, and the horizontal axis indicates the wave number $k= 2\pi /x $.

The Meszaros and sound-wave effects suppress the amplitudes of the matter fluctuations on scales less than the horizon scale $r_\mathrm{eq}$ at the equality time, while the amplitudes on scales more than $r_\mathrm{eq}$ are not affected by the Meszaros and sound-wave effects. 
As a result, the peak of the MPS is located near the wave number $k_\mathrm{eq}$, owing to $r_\mathrm{eq}$, as shown in Fig. \ref{fig_supp_1}.

The term of the radiation-dominated era depends on $\rho_\mathrm{R}$. The time evolution of the ensemble energy density of the PMF $\rho_\mathrm{MF}$ is inversely proportional to $a^4$ as $\rho_\mathrm{R}$. Therefore, considering $\rho_\mathrm{MF}$ correctly, 
the term of the radiation-dominated era becomes longer and the horizon scale at the end of this era becomes larger. As a result, the peak of MPS $k_p$ is shifted to smaller wave number (larger scale) and the amplitude of MPS is suppressed in $k>k_p$, as shown in Fig. \ref{fig_supp_1} \cite{2016PhRvD..93d3004Y}.
Since the PMF also increases the sound velocity, the sound-wave effect is boosted \cite{2014PhRvD..89j3528Y,2016PhRvD..93d3004Y} in $k>k_p$ (Fig. \ref{fig_supp_1}). 

MPS owing to the perturbative PMF was previously thought to be major effects of the PMF on the MPS at smaller scales
Considering suppressions owing to the background PMF correctly, 
MPSs from the perturbative PMF source, also, are suppressed by the Meszaros and sound-wave effects at smaller scales as shown in the bottom panels of Fig. \ref{fig_supp_1}.
Finally, in the linear regime, the suppression from the background PMF has a major effect on the MPS; however, the increase from the perturbative PMF has secondary effects on the MPS, as shown in the top panels of Fig. \ref{fig_supp_1}.
\section{Lensing potential and weak lensing in the PMF}
The WL effect depends on the lensing potential (LP) \cite{2005PhRvD..71j3010C,2006PhR...429....1L}, and the spectrum of the LP $C_\ell^{\psi}$ depends on the MPS \cite{2006PhR...429....1L}(see Appendix \ref{sec:AII}).

As mentioned in Sec. \ref{sec:III}, the peak position of the MPS depends on the Meszaros effect \cite{1974A&A....37..225M}, and the amplitudes of the MPS at scales smaller than the peak position of the MPS are suppressed by the Meszaros \cite{1974A&A....37..225M} and sound-wave effects \cite{Peebles:1980booka,Ma:1995ey,1996ApJ...471..542H,Yamamoto:1997qc}.
The background PMF effects owing to the PMF energy density contribute to the Meszaros and sound-wave effects and change the MPS features. As a result, the peak position of the MPS shifts to smaller wave numbers (larger scales), and the amplitude of the MPS is suppressed for wave numbers larger than 0.01 Mpc$^{-1}$ by the PMF (Sec. \ref{sec:III}).
From Ref. \cite{2016PhRvD..93d3004Y} and Sec. \ref{sec:III}, the perturbative PMF effects increase the amplitudes of the MPS for wave numbers larger than 0.1 Mpc$^{-1}$. The MPSs from the perturbative PMF effects are also suppressed by the background PMF effects, as shown in Sec. \ref{sec:III}. Finally, the overall effect of the PMF suppresses the MPS for wave numbers larger than 0.01 Mpc$^{-1}$.

\begin{figure}
\includegraphics[width=1.0\textwidth]{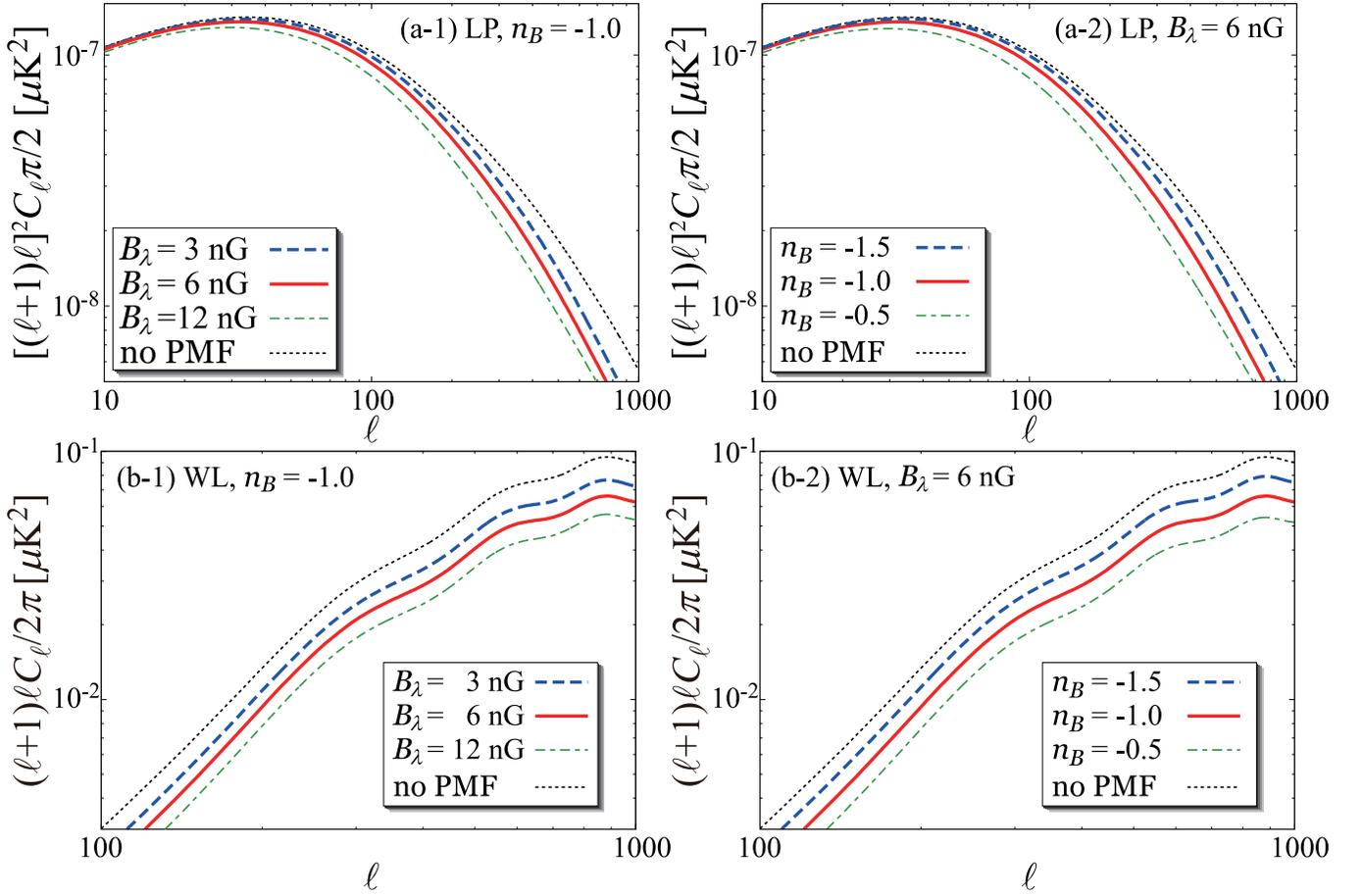}
\caption{\label{fig1}
PMF effects on the LP spectra (top panels) and CMB $B$ modes owing to WL (bottom panels). The thin dotted curves denote the theoretical result without PMF effects and based on the cosmological parameters determined by Planck 2015 (TT + lowP + lensing in Table 4 of Ref. \cite{2016A&A...594A..13P}).
The dashed, bold, and dashed-dotted curves are the theoretical results with the PMF effects, as shown by the legends in the panels.
} 
\end{figure}

\begin{figure}
\includegraphics[width=1.0\textwidth]{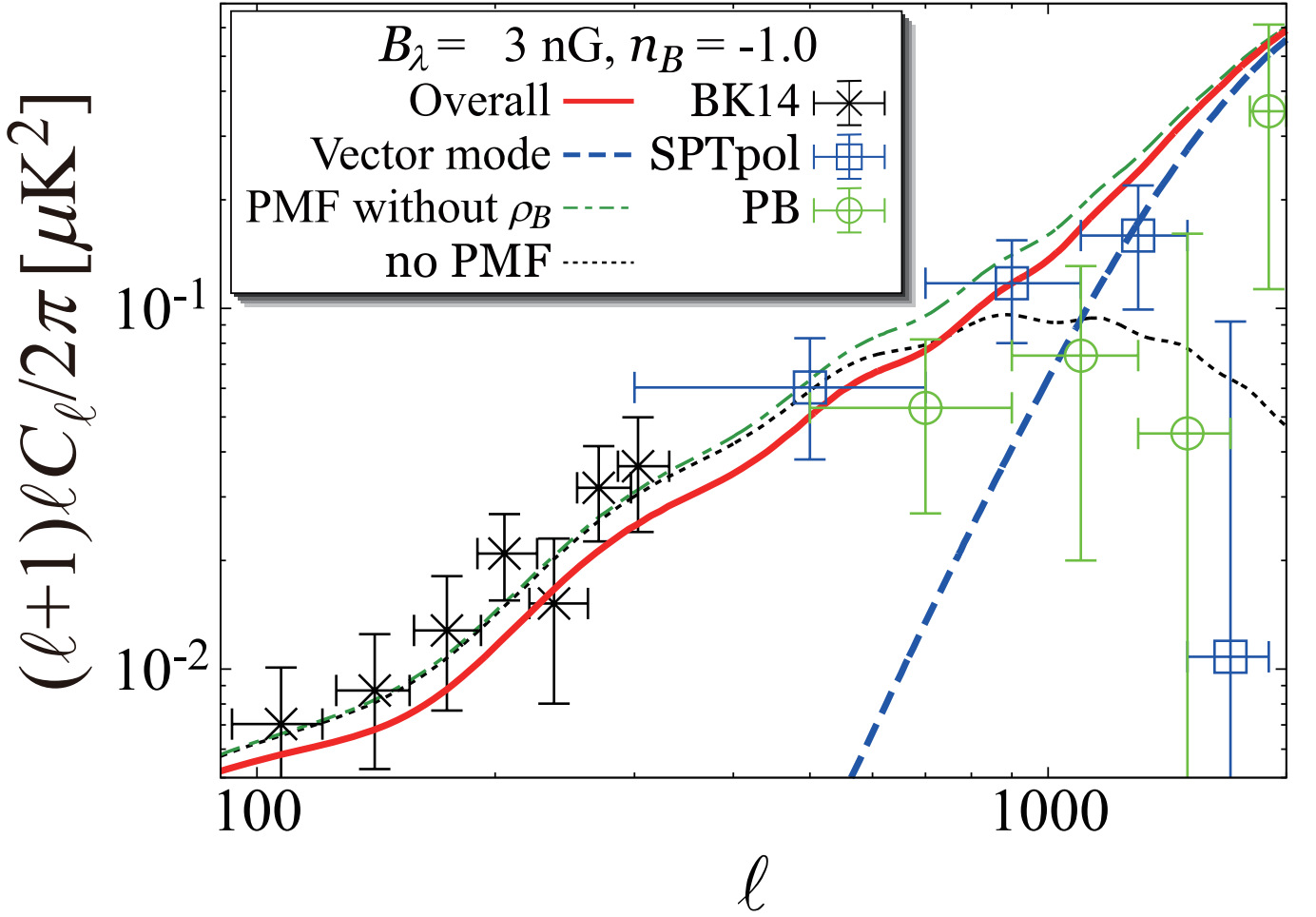}
\caption{\label{fig2}
PMF effects on the CMB $B$ mode for fixed PMF parameters $(B_\lambda,~n_\mathrm{B})~=~(3~\mathrm{nG},~-1.0)$.
The thin dotted curve denotes the theoretical result with the tensor mode and without PMF effects, and it is computed by the cosmological parameters from Planck 2015 (TT + lowP + lensing in Table 4 of Ref. \cite{2016A&A...594A..13P}).
The tensor-to-scalar ratio $r$ is 0.05.
The dashed curve is the vector mode from the PMF. The bold and dashed-dotted curves are the theoretical results from overall PMF effects with and without $\rho_B$, respectively.
The dots with error bars are the CMB $B$ mode measurements from BICEP2/Keck (BK14) \cite{2016PhRvL.116c1302B}, POLARBEAR(PB) \cite{2017ApJ...848..121T}, and SPTpol \cite{2015ApJ...807..151K}, as shown by the legends. The uncertainties of the error bars correspond to 1$\sigma$ (68.3\% confidence level).
} 
\end{figure}

From the PMF effects on the MPS, the LP spectra with the PMF are as shown in panels (a-1) and (a-2) of Fig. \ref{fig1}. 
Background PMFs shift the peak positions $\ell_\mathrm{P}$ of the LP spectra to smaller multipoles (larger scales) and suppress the amplitudes for $\ell \gtrsim \ell_\mathrm{P}$.
The WL effect on the CMB $B$ mode is due to the LP; hence, the PMFs affect the CMB $B$ mode owing to WL through the LP [panels (b-1) and (b-2) in Fig. \ref{fig1}]. 
The vertical axis of panels (a-1) and (a-2) in Fig. \ref{fig1} is $[(\ell+1)\ell]^2 C_\ell /2\pi [\mu\mathrm{K}^2]$; however, the vertical axis in panels (b-1) and (b-2) in Fig. \ref{fig1} is $(\ell+1)\ell C_\ell /2\pi [\mu\mathrm{K}^2]$. 
 To analyze the WL effect on the CMB $B$ mode owing only to the LP with the PMFs, the CMB $B$ mode spectra of vector and tensor sources are not included in panels (b-1) and (b-2) of Fig. \ref{fig1}.
If the background PMF effect for estimating the MPS is not considered, the LP spectra and CMB $B$ mode from the WL effect with the PMF in smaller multipoles remain nearly unchanged like the MPS without the background PMF (Sec. \ref{sec:III}). In fact, the overall PMF effect suppresses the PL spectra and the CMB $B$ mode owing to WL for $\ell > 100$, as shown in Figs. \ref{fig1} and \ref{fig2}.

Unlike the MPS and the LP spectrum, the CMB $B$ mode from the perturbative PMF effect has two kind modes: vector and tensor mode. The vector and tensor CM$B$ modes from the perturbative PMF effect directly increase the amplitude of the CMB $B$ mode \cite{Durrer:1999bk,Mack:2001gc,Lewis:2004ef,2010PhRvD..81b3008Y,2012PhR...517..141Y,2012PhRvD..86d3510S}. The tensor CMB $B$ mode from the perturbative PMF is much smaller than the CMB $B$ mode from the WL effect, whereas the CMB $B$ mode of the vector mode from the perturbative PMF effect corresponds to the CMB $B$ mode from the WL effect for hundreds of multipoles.
In fact, as shown in Fig. \ref{fig2} for $\ell > 800$, it directly increases the amplitude of the CMB $B$ mode spectra and dominates the spectra.
However, because the error bars of the data sets for $\ell > 800$ are large (Fig. \ref{fig2}), it is difficult to constrain the parameters of the PMF based on the observation data sets. 
In addition, the CMB $B$ mode from the perturbative PMF effect for $\ell < 400$, where the observational data points have smaller errors from BK14 \cite{2016PhRvL.116c1302B} (Fig. \ref{fig2}), is too small to effectively constrain the PMF parameters. Nevertheless, the background PMF effect suppresses the CMB $B$ mode owing to WL in the effective range of BK14 \cite{2016PhRvL.116c1302B}, as shown in Fig. \ref{fig2}. 
Thus, if we use the observational CMB $B$ mode data sets for $\ell < 400$, the PMF parameters are constrained.
\section{Degeneracies between the PMF and cosmological parameters}
Finally, we discuss the possibility of degeneracies among the PMF, matter density and baryon density parameters. The matter-to-radiation ratio determines the scale factor $a_\mathrm{eq}$ at the matter-radiation equality as follows:
\begin{eqnarray}
a_\mathrm{eq} = \frac{\rho_\mathrm{R}}{\rho_\mathrm{M}}.\label{eq:eq}
\end{eqnarray} 
From this relation and the Meszaros effect that depends on the horizon scale at $a_\mathrm{eq}$ \cite{1974A&A....37..225M}, the effect of decreasing $\rho_\mathrm{M}$ on the CMB $B$ mode owing to WL is similar to the effect of adding the background PMF to the CMB $B$ mode. 
Since a larger baryon density enforces the diffusion damping during the epoch of recombination \cite{1968ApJ...151..459S} and suppresses the amplitude of the MPS at small scales, the effect of increasing $\rho_\mathrm{b}$ on the CMB $B$ mode owing to the WL effect is similar to the effect of adding the background PMF onto the CMB $B$ mode.

\begin{figure}
\includegraphics[width=1.0\textwidth]{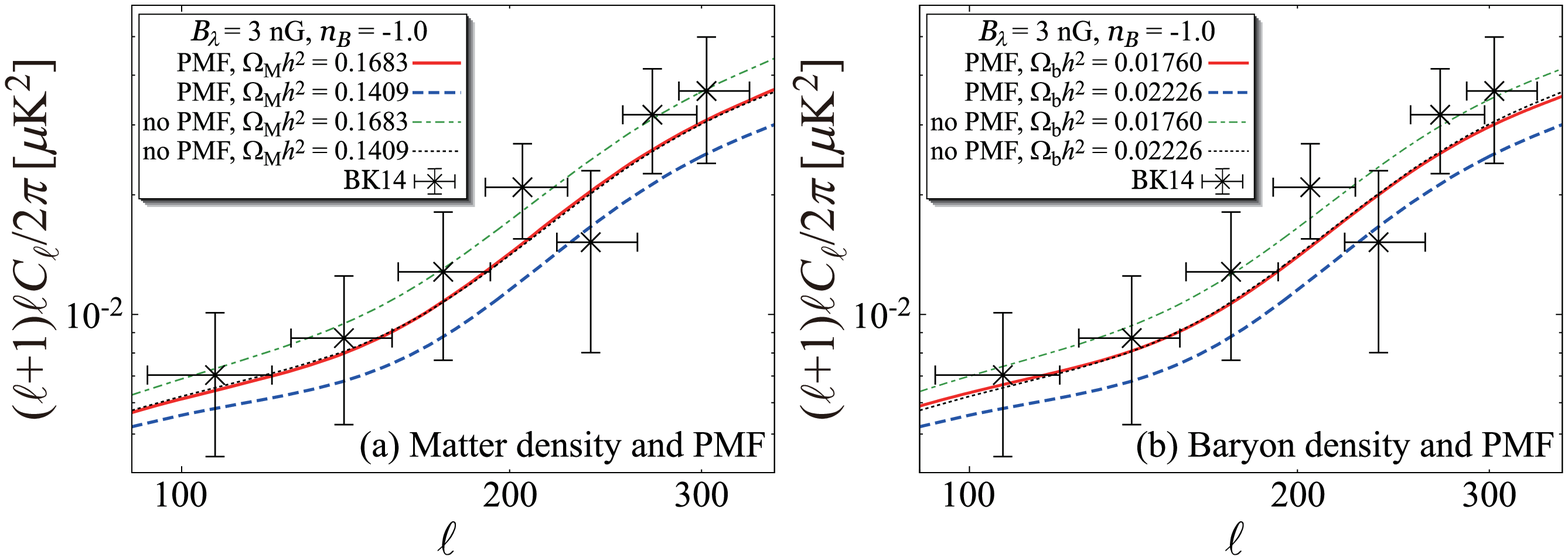}
\caption{\label{fig3}
The contribution of the PMF, matter density ($\Omega_\mathrm{M}h^2$),and baryon density ($\Omega_\mathrm{b}h^2$) parameters to the CMB $B$ mode for fixed PMF parameters $(B_\lambda,~n_\mathrm{B})~=~(3~\mathrm{nG},~-1.0)$.
The thin dotted curves denote the theoretical results with the tensor mode and without PMF effects, and they are computed by the cosmological parameters from the Planck 2015 (TT + lowP + lensing in Table 4 of Ref. \cite{2016A&A...594A..13P}).
The tensor-to-scalar ratio $r$ is 0.05.
The dots with error bars are the CMB $B$ mode measurements from BICEP2/Keck (BK14) \cite{2016PhRvL.116c1302B}.
The uncertainties of the error bars correspond to 1$\sigma$ (68.3\% confidence level).
The bold and dashed curves in panel (a) denote the theoretical results for the PMF effects of $\Omega_\mathrm{M}h^2 = 0.1683$($\Omega_\mathrm{C}h^2 = 0.1460$, $\Omega_\mathrm{b}h^2 = 0.02226$) and $\Omega_\mathrm{M}h^2 = 0.1409$($\Omega_\mathrm{C}h^2 = 0.1186$, $\Omega_\mathrm{b}h^2 = 0.02226$), respectively.
The dashed-dotted curve in panel (a) is the theoretical result without the PMF effects of $\Omega_\mathrm{M}h^2 = 0.1683$($\Omega_\mathrm{C}h^2 = 0.1460$, $\Omega_\mathrm{b}h^2 = 0.02226$). 
The CDM density parameter is fixed as $\Omega_\mathrm{C}h^2 = 0.11186$ in panel (b).
The bold and dashed curves in panel (b) denote the theoretical result with the overall PMF effects of $\Omega_\mathrm{b}h^2 = 0.01760$ and $\Omega_\mathrm{b}h^2 = 0.02226$, respectively.
The dashed-dotted curve in panel (b) is the theoretical result without the PMF effects of $\Omega_\mathrm{b}h^2 = 0.01760$. 
} 
\end{figure}

Thus, there may be degeneracies between matter and baryon density and the PMF parameters. In the absence of the PMF effect, as shown by the dashed-dotted curve in panels (a) and (b) of Fig. \ref{fig3}, the theoretical computed CMB $B$ modes of $\Omega_\mathrm{M} h^2= 0.1653$ [panel (a)] and $\Omega_\mathrm{b} h^2= 0.01760$ [panel (b)] are larger than $\Omega_\mathrm{M} h^2= 0.1409$ [panel (a)] and $\Omega_\mathrm{b} h^2= 0.02226$ [panel (b)], respectively, 
The increased CMB $B$ mode by the larger matter density parameter ($\Omega_\mathrm{M} h^2= 0.1653$) or the smaller baryon density parameter ($\Omega_\mathrm{b} h^2= 0.01760$) is modified by the PMF effect of $(B_\lambda, ~n_B) = (3~ \mathrm{nG},~ -1.0)$, as shown in panel (a) or (b) in Fig. \ref{fig3}, respectively. 
If one does not consider the background PMF effect on the CMB $B$ mode, then the PMF parameters are not constrained and the degeneracies are not correctly analyzed. Therefore, to understand the total PMF effects in the early Universe and correctly constrain the cosmological parameters including the PMF ones, we should simultaneously consider the background and perturbative PMF effects.
\section{Summary}
It has been known that the vector mode from the perturbative PMF effect increases the CMB $B$ mode for large multipoles, e.g., more than hundreds of multipoles. 
In this paper, we find that, as in the background PMF effect on the MPS, the background PMF effect suppresses the LP spectrum for a wider range of multipoles. Since the CMB $B$ mode owing to WL depends on the LP spectrum, the background PMF effect indirectly suppresses it for a wider range of multipoles. We also show that we expect to strongly constrain the PMF parameters from observational data sets at $\ell < 400$, which have relatively smaller errors, if we correctly consider the background PMF effect on the CMB $B$ mode. Finally, we discuss the possibility of non-negligible degeneracy between the PMF and matter density parameters.
The magnetic field affects the physical process at wide scale ranges in the Universe. However, studies of magnetic fields at cosmological scales are less active because it is difficult to directly observe the magnetic fields at cosmological scales and evaluate theoretical models considering magnetic fields with less observation and results. If we adopt the background PMF effects on the CMB $B$ mode and constrain the PMF parameters in the CMB $B$ mode with observations having smaller errors, we can promote the study of PMF in cosmology.

\begin{acknowledgments}
This work has been supported in part by Grants-in-Aid for Scientific Research 
(Grant No. 25871055) of the Ministry of Education, Culture, Sports,
Science and Technology of Japan.
\end{acknowledgments}

\appendix
\section{The power law PMF model\label{sec:AI}}
We adopt a PL PMF model \cite{Mack:2001gc,2014PhRvD..89j3528Y,2016PhRvD..93d3004Y}. 
The PL spectrum of the PMF on wave number $k$ for the comoving coordinates is
\begin{eqnarray}
\langle B(k)B^\ast(k)\rangle  \propto k^{n_\mathrm{B}},\label{plPMF} 
\end{eqnarray}
where $n_\mathrm{B}$ is the power law index for the PL PMF, 
and $B(k)$ is the current value(at the scale factor $a=1$).
The two-point correlation of the PL PMF is
\begin{eqnarray}
\left\langle B^{i}(\mbi{k}) {B^{j}}^*(\mbi{k}')\right\rangle 
	=	({(2\pi)^{n_\mathrm{B}+8}}/{2k_\lambda^{n_\mathrm{B}+3}})
		[{B^2_{\lambda}}/{\Gamma\left(\frac{n_\mathrm{B}+3}{2}\right)}]
		k^{n_\mathrm{B}}P^{ij}(k)\delta(\mbi{k}-\mbi{k}'), 
                k < k_\mathrm{max},
		\label{two_point1} 
\end{eqnarray}
where 
$k$ is the wave number;
$n_\mathrm{B}$ is the spectrum index of the PMF;
$B_\lambda=|\bm{B}_\lambda|$ is the comoving field strength by smoothing over a Gaussian sphere of radius $\lambda=1$ Mpc ($k_\lambda = 2\pi/\lambda$); 
$\Gamma(x)$ is the gamma function; 
$i$ and $j$ are the spatial indices and the integer numbers [$\in$ (1, 2, 3)]; 
$P^{ij}(k)=\delta^{ij}-\frac{k{}^{i}k{}^{j}}{k{}^2}$; 
$k_\mathrm{max}$ is the cutoff scale and is defined by the PMF damping \cite{Jedamzik:1996wp, Subramanian:1997gi,Mack:2001gc}; 
and 
$B^{i}(\mbi{k})$ is from the Fourier transform convention:
$B^{i}(\mbi{k}) = \int d^3 x e^{i\mbi{k}\cdot\mbi{x}} B^{i}(\mbi{x})$.
Here $B(a,\mbi{x}) = B(\mbi{x})/a^2$, where $a$ is the scale factor.
The ensemble average of the PMF energy density $\rho_\mathrm{MF}$
on the physical field is derived as follows \cite{Subramanian:1997gi,Subramanian:1998fn,2012PhRvD..86l3006Y,2014PhRvD..89j3528Y,2016PhRvD..93d3004Y}:
\begin{eqnarray}
\rho_\mathrm{MF}
= 
\frac{1}{8\pi a^4}
\frac{
B^2_\lambda
}
{
   \Gamma
   \left(
      \frac{n_\mathrm{B}+5}{2}
   \right)
}
\left[
(\lambda k_\mathrm{max})^{n_\mathrm{B}+3}
-
(\lambda k_\mathrm{min})^{n_\mathrm{B}+3}
\right],
\label{eq:BG_PL_PMF_energy_density}
\end{eqnarray}
where 
$k_\mathrm{min}$ is the minimum wave number.
We assume that a PMF is produced by some vorticity anisotropies from an inflationary source; thus, $k_\mathrm{[min]}/k_\mathrm{max}$ can be assumed to be very small. In such case, the second term of Eq. (\ref{eq:BG_PL_PMF_energy_density}) is negligible, and this equation becomes
\begin{eqnarray}
\rho_\mathrm{MF}
\sim
\frac{1}{8\pi a^4}
\frac{
B^2_\lambda
}
{
   \Gamma
   \left(
      \frac{n_\mathrm{B}+5}{2}
   \right)
}
(\lambda k_\mathrm{max})^{n_\mathrm{B}+3}.
\label{eq:BG_PL_PMF_energy_densityII}
\end{eqnarray} 
\section{The lensing potential and weak lensing\label{sec:AII}}
The power spectrum of the lensing potential is \cite{2005PhRvD..71j3010C,2006PhR...429....1L}
\begin{eqnarray}
C^{\psi}_\ell
=
16\pi 
\int\frac{dk}{k}
P_\mathrm{R}(k)
\left[
\int^{\chi_\ast}_{0}d\chi^\prime
T_\Psi(k; \eta_0~-~\chi)
j_\ell (k\chi) 
j_\ell (k\chi^\prime) 
\left(
\frac{\chi_\ast-\chi}{\chi_\ast \chi}
\right)
\right]^2
,
\label{eq_lps}
\end{eqnarray}
where
$k$ is the wave number and is defined by $k=2\pi/x$, 
$P_\mathrm{R}(k)$ is the primary power spectrum with or without the background PMF, 
$\chi_\ast$ is the comoving distance from the CMB photons to the observer, 
$\chi$ is the comoving distance from the potentials to the observer, 
$T_\Psi$ is the transfer function,
$j_\ell (X)$ is the spherical Bessel function 
defined by $j_\ell(X)=(\pi/2X)^{1/2}J_{\ell+1/2}(X)$, and $J_\ell(X)$ is the standard Bessel functions.
This power spectrum is related to the MPS \cite{2006PhR...429....1L}.

The $B$ mode correlation functions from the WL effect in the curved sky case are \cite{2006PhR...429....1L}
\begin{eqnarray}
\tilde{\xi}_{+} 
\sim \sum_{\ell}
\frac{2\ell + 1}{4\pi}
\left(
C^{E}_\ell + C^{B}_\ell
\right)
\exp
\left(-\ell(\ell+1)\frac{\sigma^2}{2}
\right)
\nonumber\\
\times
\left[
d^\ell_{22} + \frac{1}{2}\ell(\ell+1)C_\mathrm{gl,2}d^\ell_{31} +~\cdots~
\right]
\\
\tilde{\xi}_{-} 
\sim \sum_{\ell}
\frac{2\ell + 1}{4\pi}
\left(
C^{E}_\ell - C^{B}_\ell
\right)
\exp
\left(-\ell(\ell+1)\frac{\sigma^2}{2}
\right)
\nonumber\\
\times
\left[
d^\ell_{2-2} + \frac{1}{4}\ell(\ell+1)C_\mathrm{gl,2}(d^\ell_{1-1}+d^\ell_{3-3}) +~\cdots~
\right]
\\
\tilde{C}^B_\ell = 
\pi
\int^{1}_{-1}
\left(
\tilde{\xi}_{+}(\beta)d^{\ell}_{22}(\beta)
-
\tilde{\xi}_{-}(\beta)d^{\ell}_{2-2}(\beta)
\right)
\mathrm{d}\cos\beta,
\end{eqnarray}
where 
$C_\ell^{E}$ and $C_\ell^{B}$ are the unlensed $E$ and $B$ mode power spectra, and 
$d^{\ell}_{mm\prime}$ is the reduced Wigner functions and is defined by
$d^{\ell}_{mm\prime}(\beta) \equiv D^\ell_{mm^\prime}(0,\beta,0)$.
Here, also, 
$\D C_\mathrm{gl}(\beta) = \sum_{\ell}\frac{2\ell+1}{4\pi}\ell(\ell+1)C^\psi_\ell d^\ell_{11}(\beta)$,
$\D C_\mathrm{gl,2}(\beta) = \sum_{\ell}\frac{2\ell+1}{4\pi}\ell(\ell+1)C^\psi_\ell d^\ell_{-11}(\beta)$, and 
$\sigma(\beta) \equiv C_\mathrm{gl}(0) - C_\mathrm{gl}(\beta)$.

\newpage
\bibliographystyle{apsrev}

\begin{thebibliography}{63}
\expandafter\ifx\csname natexlab\endcsname\relax\def\natexlab#1{#1}\fi
\expandafter\ifx\csname bibnamefont\endcsname\relax
  \def\bibnamefont#1{#1}\fi
\expandafter\ifx\csname bibfnamefont\endcsname\relax
  \def\bibfnamefont#1{#1}\fi
\expandafter\ifx\csname citenamefont\endcsname\relax
  \def\citenamefont#1{#1}\fi
\expandafter\ifx\csname url\endcsname\relax
  \def\url#1{\texttt{#1}}\fi
\expandafter\ifx\csname urlprefix\endcsname\relax\def\urlprefix{URL }\fi
\providecommand{\bibinfo}[2]{#2}
\providecommand{\eprint}[2][]{\url{#2}}

\bibitem[{\citenamefont{{Louis} et~al.}(2017)\citenamefont{{Louis}, {Grace},
  {Hasselfield}, {Lungu}, {Maurin}, {Addison}, {Ade}, {Aiola}, {Allison},
  {Amiri} et~al.}}]{2017JCAP...06..031L}
\bibinfo{author}{\bibfnamefont{T.}~\bibnamefont{{Louis}}},
  \bibinfo{author}{\bibfnamefont{E.}~\bibnamefont{{Grace}}},
  \bibinfo{author}{\bibfnamefont{M.}~\bibnamefont{{Hasselfield}}},
  \bibinfo{author}{\bibfnamefont{M.}~\bibnamefont{{Lungu}}},
  \bibinfo{author}{\bibfnamefont{L.}~\bibnamefont{{Maurin}}},
  \bibinfo{author}{\bibfnamefont{G.~E.} \bibnamefont{{Addison}}},
  \bibinfo{author}{\bibfnamefont{P.~A.~R.} \bibnamefont{{Ade}}},
  \bibinfo{author}{\bibfnamefont{S.}~\bibnamefont{{Aiola}}},
  \bibinfo{author}{\bibfnamefont{R.}~\bibnamefont{{Allison}}},
  \bibinfo{author}{\bibfnamefont{M.}~\bibnamefont{{Amiri}}},
  \bibnamefont{$et~al.$}, \bibinfo{journal}{\jcap} 
  \bibinfo{volume}{6} (\bibinfo{year}{2017}) \bibinfo{eid}{031} .

\bibitem[{\citenamefont{{Keisler} et~al.}(2015)\citenamefont{{Keisler},
  {Hoover}, {Harrington}, {Henning}, {Ade}, {Aird}, {Austermann}, {Beall},
  {Bender}, {Benson} et~al.}}]{2015ApJ...807..151K}
\bibinfo{author}{\bibfnamefont{R.}~\bibnamefont{{Keisler}}},
  \bibinfo{author}{\bibfnamefont{S.}~\bibnamefont{{Hoover}}},
  \bibinfo{author}{\bibfnamefont{N.}~\bibnamefont{{Harrington}}},
  \bibinfo{author}{\bibfnamefont{J.~W.} \bibnamefont{{Henning}}},
  \bibinfo{author}{\bibfnamefont{P.~A.~R.} \bibnamefont{{Ade}}},
  \bibinfo{author}{\bibfnamefont{K.~A.} \bibnamefont{{Aird}}},
  \bibinfo{author}{\bibfnamefont{J.~E.} \bibnamefont{{Austermann}}},
  \bibinfo{author}{\bibfnamefont{J.~A.} \bibnamefont{{Beall}}},
  \bibinfo{author}{\bibfnamefont{A.~N.} \bibnamefont{{Bender}}},
  \bibinfo{author}{\bibfnamefont{B.~A.} \bibnamefont{{Benson}}},
  \bibnamefont{$et~al.$}, \bibinfo{journal}{\apj} \textbf{\bibinfo{volume}{807}},
  \bibinfo{eid}{151} (\bibinfo{year}{2015}).

\bibitem[{\citenamefont{{The POLARBEAR Collaboration}
  et~al.}(2017)\citenamefont{{The POLARBEAR Collaboration}, {Ade}, {Aguilar},
  {Akiba}, {Arnold}, {Baccigalupi}, {Barron}, {Beck}, {Bianchini}, {Boettger}
  et~al.}}]{2017ApJ...848..121T}
  \bibinfo{author}{\bibfnamefont{P.~A.~R.} \bibnamefont{{Ade}}},
  \bibinfo{author}{\bibfnamefont{M.}~\bibnamefont{{Aguilar}}},
  \bibinfo{author}{\bibfnamefont{Y.}~\bibnamefont{{Akiba}}},
  \bibinfo{author}{\bibfnamefont{K.}~\bibnamefont{{Arnold}}},
  \bibinfo{author}{\bibfnamefont{C.}~\bibnamefont{{Baccigalupi}}},
  \bibinfo{author}{\bibfnamefont{D.}~\bibnamefont{{Barron}}},
  \bibinfo{author}{\bibfnamefont{D.}~\bibnamefont{{Beck}}},
  \bibinfo{author}{\bibfnamefont{F.}~\bibnamefont{{Bianchini}}},
  \bibinfo{author}{\bibfnamefont{D.}~\bibnamefont{{Boettger}}},
  \bibnamefont{$et~al.$}(\bibinfo{author}{\bibnamefont{{The POLARBEAR Collaboration}}}),
  \bibinfo{journal}{\apj} \textbf{\bibinfo{volume}{848}},
  \bibinfo{eid}{121} (\bibinfo{year}{2017}).

\bibitem[{\citenamefont{{BICEP2 Collaboration}
  et~al.}(2016)\citenamefont{{BICEP2 Collaboration}, {Keck Array
  Collaboration}, {Ade}, {Ahmed}, {Aikin}, {Alexander}, {Barkats}, {Benton},
  {Bischoff}, {Bock} et~al.}}]{2016PhRvL.116c1302B}
  \bibinfo{author}{\bibfnamefont{P.~A.~R.} \bibnamefont{{Ade}}},
  \bibinfo{author}{\bibfnamefont{Z.}~\bibnamefont{{Ahmed}}},
  \bibinfo{author}{\bibfnamefont{R.~W.} \bibnamefont{{Aikin}}},
  \bibinfo{author}{\bibfnamefont{K.~D.} \bibnamefont{{Alexander}}},
  \bibinfo{author}{\bibfnamefont{D.}~\bibnamefont{{Barkats}}},
  \bibinfo{author}{\bibfnamefont{S.~J.} \bibnamefont{{Benton}}},
  \bibinfo{author}{\bibfnamefont{C.~A.} \bibnamefont{{Bischoff}}},
  \bibinfo{author}{\bibfnamefont{J.~J.} \bibnamefont{{Bock}}},
  \bibnamefont{$et~al.$}(\bibinfo{author}{\bibnamefont{{BICEP2 Collaboration}}}, \bibinfo{author}{\bibnamefont{{Keck Array Collaboration}}}),
  \bibinfo{journal}{\prl} \textbf{\bibinfo{volume}{116}},
  \bibinfo{eid}{031302} (\bibinfo{year}{2016}).

\bibitem[{\citenamefont{{Govoni} and {Feretti}}(2004)}]{2004IJMPD..13.1549G}
\bibinfo{author}{\bibfnamefont{F.}~\bibnamefont{{Govoni}}} \bibnamefont{and}
  \bibinfo{author}{\bibfnamefont{L.}~\bibnamefont{{Feretti}}},
  \bibinfo{journal}{International Journal of Modern Physics D}
  \textbf{\bibinfo{volume}{13}}, \bibinfo{pages}{1549} (\bibinfo{year}{2004}).

\bibitem[{\citenamefont{Wolfe et~al.}(1992)\citenamefont{Wolfe, Lanzetta, and
  Oren}}]{Wolfe:1992ab}
\bibinfo{author}{\bibfnamefont{A.~M.} \bibnamefont{Wolfe}},
  \bibinfo{author}{\bibfnamefont{K.~M.} \bibnamefont{Lanzetta}},
  \bibnamefont{and} \bibinfo{author}{\bibfnamefont{A.~L.} \bibnamefont{Oren}},
  \bibinfo{journal}{\apj} \textbf{\bibinfo{volume}{388}}, \bibinfo{pages}{17}
  (\bibinfo{year}{1992}).

\bibitem[{\citenamefont{Clarke et~al.}(2001)\citenamefont{Clarke, Kronberg, and
  Boehringer}}]{Clarke:2000bz}
\bibinfo{author}{\bibfnamefont{T.~E.} \bibnamefont{Clarke}},
  \bibinfo{author}{\bibfnamefont{P.~P.} \bibnamefont{Kronberg}},
  \bibnamefont{and}
  \bibinfo{author}{\bibfnamefont{H.}~\bibnamefont{Boehringer}},
  \bibinfo{journal}{Astrophys. J.} \textbf{\bibinfo{volume}{547}},
  \bibinfo{pages}{L111} (\bibinfo{year}{2001}).

\bibitem[{\citenamefont{Xu et~al.}(2006)\citenamefont{Xu, Kronberg, Habib, and
  Dufton}}]{Xu:2005rb}
\bibinfo{author}{\bibfnamefont{Y.}~\bibnamefont{Xu}},
  \bibinfo{author}{\bibfnamefont{P.~P.} \bibnamefont{Kronberg}},
  \bibinfo{author}{\bibfnamefont{S.}~\bibnamefont{Habib}}, \bibnamefont{and}
  \bibinfo{author}{\bibfnamefont{Q.~W.} \bibnamefont{Dufton}},
  \bibinfo{journal}{Astrophys. J.} \textbf{\bibinfo{volume}{637}},
  \bibinfo{pages}{19} (\bibinfo{year}{2006}).

\bibitem[{\citenamefont{{Neronov} and {Vovk}}(2010)}]{2010Sci...328...73N}
\bibinfo{author}{\bibfnamefont{A.}~\bibnamefont{{Neronov}}} \bibnamefont{and}
  \bibinfo{author}{\bibfnamefont{I.}~\bibnamefont{{Vovk}}},
  \bibinfo{journal}{Science} \textbf{\bibinfo{volume}{328}},
  \bibinfo{pages}{73} (\bibinfo{year}{2010}).

\bibitem[{\citenamefont{{Vovk} et~al.}(2012)\citenamefont{{Vovk}, {Taylor},
  {Semikoz}, and {Neronov}}}]{2012ApJ...747L..14V}
\bibinfo{author}{\bibfnamefont{I.}~\bibnamefont{{Vovk}}},
  \bibinfo{author}{\bibfnamefont{A.~M.} \bibnamefont{{Taylor}}},
  \bibinfo{author}{\bibfnamefont{D.}~\bibnamefont{{Semikoz}}},
  \bibnamefont{and}
  \bibinfo{author}{\bibfnamefont{A.}~\bibnamefont{{Neronov}}},
  \bibinfo{journal}{\apjl} \textbf{\bibinfo{volume}{747}}, \bibinfo{eid}{L14}
  (\bibinfo{year}{2012}).

\bibitem[{\citenamefont{{Feretti} et~al.}(2012)\citenamefont{{Feretti},
  {Giovannini}, {Govoni}, and {Murgia}}}]{2012A&ARv..20...54F}
\bibinfo{author}{\bibfnamefont{L.}~\bibnamefont{{Feretti}}},
  \bibinfo{author}{\bibfnamefont{G.}~\bibnamefont{{Giovannini}}},
  \bibinfo{author}{\bibfnamefont{F.}~\bibnamefont{{Govoni}}}, \bibnamefont{and}
  \bibinfo{author}{\bibfnamefont{M.}~\bibnamefont{{Murgia}}},
  \bibinfo{journal}{\aapr} \textbf{\bibinfo{volume}{20}}, \bibinfo{pages}{54}
  (\bibinfo{year}{2012}).

\bibitem[{\citenamefont{Grasso and Rubinstein}(2001)}]{Grasso:2000wj}
\bibinfo{author}{\bibfnamefont{D.}~\bibnamefont{Grasso}} \bibnamefont{and}
  \bibinfo{author}{\bibfnamefont{H.~R.} \bibnamefont{Rubinstein}},
  \bibinfo{journal}{\physrep} \textbf{\bibinfo{volume}{348}},
  \bibinfo{pages}{163} (\bibinfo{year}{2001}).

\bibitem[{\citenamefont{{Yamazaki}
  et~al.}(2010{\natexlab{a}})\citenamefont{{Yamazaki}, {Ichiki}, {Kajino}, and
  {Mathews}}}]{2010AdAst2010E..80Y}
\bibinfo{author}{\bibfnamefont{D.~G.} \bibnamefont{{Yamazaki}}},
  \bibinfo{author}{\bibfnamefont{K.}~\bibnamefont{{Ichiki}}},
  \bibinfo{author}{\bibfnamefont{T.}~\bibnamefont{{Kajino}}}, \bibnamefont{and}
  \bibinfo{author}{\bibfnamefont{G.~J.} \bibnamefont{{Mathews}}},
  \bibinfo{journal}{\advast} \textbf{\bibinfo{volume}{2010}},
  \bibinfo{eid}{586590} (\bibinfo{year}{2010}{\natexlab{a}}).

\bibitem[{\citenamefont{{Kandus} et~al.}(2011)\citenamefont{{Kandus}, {Kunze},
  and {Tsagas}}}]{2011PhR...505....1K}
\bibinfo{author}{\bibfnamefont{A.}~\bibnamefont{{Kandus}}},
  \bibinfo{author}{\bibfnamefont{K.~E.} \bibnamefont{{Kunze}}},
  \bibnamefont{and} \bibinfo{author}{\bibfnamefont{C.~G.}
  \bibnamefont{{Tsagas}}}, \bibinfo{journal}{\physrep}
  \textbf{\bibinfo{volume}{505}}, \bibinfo{pages}{1} (\bibinfo{year}{2011}).

\bibitem[{\citenamefont{{Yamazaki} et~al.}(2012)\citenamefont{{Yamazaki},
  {Kajino}, {Mathews}, and {Ichiki}}}]{2012PhR...517..141Y}
\bibinfo{author}{\bibfnamefont{D.~G.} \bibnamefont{{Yamazaki}}},
  \bibinfo{author}{\bibfnamefont{T.}~\bibnamefont{{Kajino}}},
  \bibinfo{author}{\bibfnamefont{G.~J.} \bibnamefont{{Mathews}}},
  \bibnamefont{and} \bibinfo{author}{\bibfnamefont{K.}~\bibnamefont{{Ichiki}}},
  \bibinfo{journal}{\physrep} \textbf{\bibinfo{volume}{517}},
  \bibinfo{pages}{141} (\bibinfo{year}{2012}).

\bibitem[{\citenamefont{{Yamazaki} et~al.}(2006)\citenamefont{{Yamazaki},
  {Ichiki}, {Umezu}, and {Hanayama}}}]{2006PhRvD..74l3518Y}
\bibinfo{author}{\bibfnamefont{D.~G.} \bibnamefont{{Yamazaki}}},
  \bibinfo{author}{\bibfnamefont{K.}~\bibnamefont{{Ichiki}}},
  \bibinfo{author}{\bibfnamefont{K.-I.} \bibnamefont{{Umezu}}},
  \bibnamefont{and}
  \bibinfo{author}{\bibfnamefont{H.}~\bibnamefont{{Hanayama}}},
  \bibinfo{journal}{\prd} \textbf{\bibinfo{volume}{74}},
  \bibinfo{pages}{123518} (\bibinfo{year}{2006}).

\bibitem[{\citenamefont{{Yamazaki} et~al.}(2008)\citenamefont{{Yamazaki},
  {Ichiki}, {Kajino}, and {Mathews}}}]{2008PhRvD..77d3005Y}
\bibinfo{author}{\bibfnamefont{D.~G.} \bibnamefont{{Yamazaki}}},
  \bibinfo{author}{\bibfnamefont{K.}~\bibnamefont{{Ichiki}}},
  \bibinfo{author}{\bibfnamefont{T.}~\bibnamefont{{Kajino}}}, \bibnamefont{and}
  \bibinfo{author}{\bibfnamefont{G.~J.} \bibnamefont{{Mathews}}},
  \bibinfo{journal}{Phys. Rev.} \textbf{\bibinfo{volume}{D 77}},
  \bibinfo{pages}{043005} (\bibinfo{year}{2008}).

\bibitem[{\citenamefont{{Finelli} et~al.}(2008)\citenamefont{{Finelli}, {Paci},
  and {Paoletti}}}]{2008PhRvD..78b3510F}
\bibinfo{author}{\bibfnamefont{F.}~\bibnamefont{{Finelli}}},
  \bibinfo{author}{\bibfnamefont{F.}~\bibnamefont{{Paci}}}, \bibnamefont{and}
  \bibinfo{author}{\bibfnamefont{D.}~\bibnamefont{{Paoletti}}},
  \bibinfo{journal}{\prd} \textbf{\bibinfo{volume}{78}}, \bibinfo{eid}{023510}
  (\bibinfo{year}{2008}), \eprint{0803.1246}.

\bibitem[{\citenamefont{{Yamazaki}
  et~al.}(2010{\natexlab{b}})\citenamefont{{Yamazaki}, {Ichiki}, {Kajino}, and
  {Mathews}}}]{2010PhRvD..81b3008Y}
\bibinfo{author}{\bibfnamefont{D.~G.} \bibnamefont{{Yamazaki}}},
  \bibinfo{author}{\bibfnamefont{K.}~\bibnamefont{{Ichiki}}},
  \bibinfo{author}{\bibfnamefont{T.}~\bibnamefont{{Kajino}}}, \bibnamefont{and}
  \bibinfo{author}{\bibfnamefont{G.~J.} \bibnamefont{{Mathews}}},
  \bibinfo{journal}{Phys. Rev.} \textbf{\bibinfo{volume}{D 81}},
  \bibinfo{pages}{023008} (\bibinfo{year}{2010}{\natexlab{b}}).

\bibitem[{\citenamefont{Durrer et~al.}(2000)\citenamefont{Durrer, Ferreira, and
  Kahniashvili}}]{Durrer:1999bk}
\bibinfo{author}{\bibfnamefont{R.}~\bibnamefont{Durrer}},
  \bibinfo{author}{\bibfnamefont{P.~G.} \bibnamefont{Ferreira}},
  \bibnamefont{and}
  \bibinfo{author}{\bibfnamefont{T.}~\bibnamefont{Kahniashvili}},
  \bibinfo{journal}{Phys. Rev.} \textbf{\bibinfo{volume}{D 61}},
  \bibinfo{pages}{043001} (\bibinfo{year}{2000}).

\bibitem[{\citenamefont{Mack et~al.}(2002)\citenamefont{Mack, Kahniashvili, and
  Kosowsky}}]{Mack:2001gc}
\bibinfo{author}{\bibfnamefont{A.}~\bibnamefont{Mack}},
  \bibinfo{author}{\bibfnamefont{T.}~\bibnamefont{Kahniashvili}},
  \bibnamefont{and} \bibinfo{author}{\bibfnamefont{A.}~\bibnamefont{Kosowsky}},
  \bibinfo{journal}{Phys. Rev.} \textbf{\bibinfo{volume}{D 65}},
  \bibinfo{pages}{123004} (\bibinfo{year}{2002}).

\bibitem[{\citenamefont{Kosowsky et~al.}(2005)\citenamefont{Kosowsky,
  Kahniashvili, Lavrelashvili, and Ratra}}]{Kosowsky:2004zh}
\bibinfo{author}{\bibfnamefont{A.}~\bibnamefont{Kosowsky}},
  \bibinfo{author}{\bibfnamefont{T.}~\bibnamefont{Kahniashvili}},
  \bibinfo{author}{\bibfnamefont{G.}~\bibnamefont{Lavrelashvili}},
  \bibnamefont{and} \bibinfo{author}{\bibfnamefont{B.}~\bibnamefont{Ratra}},
  \bibinfo{journal}{Phys. Rev.} \textbf{\bibinfo{volume}{D 71}},
  \bibinfo{pages}{043006} (\bibinfo{year}{2005}).

\bibitem[{\citenamefont{Kahniashvili and Ratra}(2005)}]{Kahniashvili:2005xe}
\bibinfo{author}{\bibfnamefont{T.}~\bibnamefont{Kahniashvili}}
  \bibnamefont{and} \bibinfo{author}{\bibfnamefont{B.}~\bibnamefont{Ratra}},
  \bibinfo{journal}{Phys. Rev.} \textbf{\bibinfo{volume}{D 71}},
  \bibinfo{pages}{103006} (\bibinfo{year}{2005}).

\bibitem[{\citenamefont{Dolgov}(2005)}]{Dolgov:2005ti}
\bibinfo{author}{\bibfnamefont{A.~D.} \bibnamefont{Dolgov}},
  \eprint{arXiv:astro-ph/0503447}.

\bibitem[{\citenamefont{Giovannini and
  Kunze}(2008{\natexlab{a}})}]{Giovannini:2007qn}
\bibinfo{author}{\bibfnamefont{M.}~\bibnamefont{Giovannini}} \bibnamefont{and}
  \bibinfo{author}{\bibfnamefont{K.~E.} \bibnamefont{Kunze}},
  \bibinfo{journal}{Phys. Rev.} \textbf{\bibinfo{volume}{D77}},
  \bibinfo{pages}{063003} (\bibinfo{year}{2008}{\natexlab{a}}).

\bibitem[{\citenamefont{Paoletti et~al.}(2009)\citenamefont{Paoletti, Finelli,
  and Paci}}]{Paoletti:2008ck}
\bibinfo{author}{\bibfnamefont{D.}~\bibnamefont{Paoletti}},
  \bibinfo{author}{\bibfnamefont{F.}~\bibnamefont{Finelli}}, \bibnamefont{and}
  \bibinfo{author}{\bibfnamefont{F.}~\bibnamefont{Paci}},
  \bibinfo{journal}{\mnras} \textbf{\bibinfo{volume}{396}},
  \bibinfo{pages}{523} (\bibinfo{year}{2009}).

\bibitem[{\citenamefont{Sethi et~al.}(2008)\citenamefont{Sethi, Nath, and
  Subramanian}}]{Sethi:2008eq}
\bibinfo{author}{\bibfnamefont{S.~K.} \bibnamefont{Sethi}},
  \bibinfo{author}{\bibfnamefont{B.~B.} \bibnamefont{Nath}}, \bibnamefont{and}
  \bibinfo{author}{\bibfnamefont{K.}~\bibnamefont{Subramanian}},
  \bibinfo{journal}{\mnras} \textbf{\bibinfo{volume}{387}},
  \bibinfo{pages}{1589} (\bibinfo{year}{2008}).

\bibitem[{\citenamefont{Kojima et~al.}(2008)\citenamefont{Kojima, Ichiki,
  Yamazaki, Kajino, and Mathews}}]{Kojima:2008rf}
\bibinfo{author}{\bibfnamefont{K.}~\bibnamefont{Kojima}},
  \bibinfo{author}{\bibfnamefont{K.}~\bibnamefont{Ichiki}},
  \bibinfo{author}{\bibfnamefont{D.~G.} \bibnamefont{Yamazaki}},
  \bibinfo{author}{\bibfnamefont{T.}~\bibnamefont{Kajino}}, \bibnamefont{and}
  \bibinfo{author}{\bibfnamefont{G.~J.} \bibnamefont{Mathews}},
  \bibinfo{journal}{Phys. Rev.} \textbf{\bibinfo{volume}{D78}},
  \bibinfo{pages}{045010} (\bibinfo{year}{2008}).

\bibitem[{\citenamefont{{Kahniashvili}
  et~al.}(2008)\citenamefont{{Kahniashvili}, {Lavrelashvili}, and
  {Ratra}}}]{2008PhRvD..78f3012K}
\bibinfo{author}{\bibfnamefont{T.}~\bibnamefont{{Kahniashvili}}},
  \bibinfo{author}{\bibfnamefont{G.}~\bibnamefont{{Lavrelashvili}}},
  \bibnamefont{and} \bibinfo{author}{\bibfnamefont{B.}~\bibnamefont{{Ratra}}},
  \bibinfo{journal}{Phys. Rev.} \textbf{\bibinfo{volume}{D78}},
  \bibinfo{pages}{063012} (\bibinfo{year}{2008}).

\bibitem[{\citenamefont{Giovannini and
  Kunze}(2008{\natexlab{b}})}]{Giovannini:2008aa}
\bibinfo{author}{\bibfnamefont{M.}~\bibnamefont{Giovannini}} \bibnamefont{and}
  \bibinfo{author}{\bibfnamefont{K.~E.} \bibnamefont{Kunze}},
  \bibinfo{journal}{Phys. Rev.} \textbf{\bibinfo{volume}{D78}},
  \bibinfo{pages}{023010} (\bibinfo{year}{2008}{\natexlab{b}}),
  \eprint{0804.3380}.

\bibitem[{\citenamefont{Shaw and Lewis}(2010)}]{Shaw:2009nf}
\bibinfo{author}{\bibfnamefont{J.~R.} \bibnamefont{Shaw}} \bibnamefont{and}
  \bibinfo{author}{\bibfnamefont{A.}~\bibnamefont{Lewis}},
  \bibinfo{journal}{Phys. Rev.} \textbf{\bibinfo{volume}{D 81}},
  \bibinfo{pages}{043517} (\bibinfo{year}{2010}).

\bibitem[{\citenamefont{{Yamazaki} et~al.}(2011)\citenamefont{{Yamazaki},
  {Ichiki}, and {Takahashi}}}]{2011PhRvD..84l3006Y}
\bibinfo{author}{\bibfnamefont{D.~G.} \bibnamefont{{Yamazaki}}},
  \bibinfo{author}{\bibfnamefont{K.}~\bibnamefont{{Ichiki}}}, \bibnamefont{and}
  \bibinfo{author}{\bibfnamefont{K.}~\bibnamefont{{Takahashi}}},
  \bibinfo{journal}{\prd} \textbf{\bibinfo{volume}{84}}, \bibinfo{eid}{123006}
  (\bibinfo{year}{2011}).

\bibitem[{\citenamefont{Yamazaki et~al.}(2005)\citenamefont{Yamazaki, Ichiki,
  and Kajino}}]{Yamazaki:2004vq}
\bibinfo{author}{\bibfnamefont{D.~G.} \bibnamefont{Yamazaki}},
  \bibinfo{author}{\bibfnamefont{K.}~\bibnamefont{Ichiki}}, \bibnamefont{and}
  \bibinfo{author}{\bibfnamefont{T.}~\bibnamefont{Kajino}},
  \bibinfo{journal}{Astrophys. J.} \textbf{\bibinfo{volume}{625}},
  \bibinfo{pages}{L1} (\bibinfo{year}{2005}).

\bibitem[{\citenamefont{Lewis}(2004)}]{Lewis:2004ef}
\bibinfo{author}{\bibfnamefont{A.}~\bibnamefont{Lewis}},
  \bibinfo{journal}{Phys. Rev.} \textbf{\bibinfo{volume}{D 70}},
  \bibinfo{pages}{043011} (\bibinfo{year}{2004}).

\bibitem[{\citenamefont{Yamazaki et~al.}(2006)\citenamefont{Yamazaki, Ichiki,
  Kajino, and Mathews}}]{Yamazaki:2006bq}
\bibinfo{author}{\bibfnamefont{D.~G.} \bibnamefont{Yamazaki}},
  \bibinfo{author}{\bibfnamefont{K.}~\bibnamefont{Ichiki}},
  \bibinfo{author}{\bibfnamefont{T.}~\bibnamefont{Kajino}}, \bibnamefont{and}
  \bibinfo{author}{\bibfnamefont{G.~J.} \bibnamefont{Mathews}},
  \bibinfo{journal}{Astrophys. J.} \textbf{\bibinfo{volume}{646}},
  \bibinfo{pages}{719} (\bibinfo{year}{2006}).

\bibitem[{\citenamefont{Yamazaki et~al.}(2008)\citenamefont{Yamazaki, Ichiki,
  Kajino, and Mathews}}]{Yamazaki:2008bb}
\bibinfo{author}{\bibfnamefont{D.~G.} \bibnamefont{Yamazaki}},
  \bibinfo{author}{\bibfnamefont{K.}~\bibnamefont{Ichiki}},
  \bibinfo{author}{\bibfnamefont{T.}~\bibnamefont{Kajino}}, \bibnamefont{and}
  \bibinfo{author}{\bibfnamefont{G.~J.} \bibnamefont{Mathews}},
  \bibinfo{journal}{Phys. Rev.} \textbf{\bibinfo{volume}{D 78}},
  \bibinfo{pages}{123001} (\bibinfo{year}{2008}).

\bibitem[{\citenamefont{Kahniashvili et~al.}(2009)\citenamefont{Kahniashvili,
  Maravin, and Kosowsky}}]{Kahniashvili:2008hx}
\bibinfo{author}{\bibfnamefont{T.}~\bibnamefont{Kahniashvili}},
  \bibinfo{author}{\bibfnamefont{Y.}~\bibnamefont{Maravin}}, \bibnamefont{and}
  \bibinfo{author}{\bibfnamefont{A.}~\bibnamefont{Kosowsky}},
  \bibinfo{journal}{Phys. Rev.} \textbf{\bibinfo{volume}{D80}},
  \bibinfo{pages}{023009} (\bibinfo{year}{2009}), \eprint{0806.1876}.

\bibitem[{\citenamefont{{Kahniashvili}
  et~al.}(2010)\citenamefont{{Kahniashvili}, {Tevzadze}, {Sethi}, {Pandey}, and
  {Ratra}}}]{2010PhRvD..82h3005K}
\bibinfo{author}{\bibfnamefont{T.}~\bibnamefont{{Kahniashvili}}},
  \bibinfo{author}{\bibfnamefont{A.~G.} \bibnamefont{{Tevzadze}}},
  \bibinfo{author}{\bibfnamefont{S.~K.} \bibnamefont{{Sethi}}},
  \bibinfo{author}{\bibfnamefont{K.}~\bibnamefont{{Pandey}}}, \bibnamefont{and}
  \bibinfo{author}{\bibfnamefont{B.}~\bibnamefont{{Ratra}}},
  \bibinfo{journal}{\prd} \textbf{\bibinfo{volume}{82}}, \bibinfo{eid}{083005}
  (\bibinfo{year}{2010}).

\bibitem[{\citenamefont{{Paoletti} and {Finelli}}(2011)}]{2011PhRvD..83l3533P}
\bibinfo{author}{\bibfnamefont{D.}~\bibnamefont{{Paoletti}}} \bibnamefont{and}
  \bibinfo{author}{\bibfnamefont{F.}~\bibnamefont{{Finelli}}},
  \bibinfo{journal}{\prd} \textbf{\bibinfo{volume}{83}},
  \bibinfo{pages}{123533} (\bibinfo{year}{2011}).

\bibitem[{\citenamefont{{Shaw} and {Lewis}}(2012)}]{2012PhRvD..86d3510S}
\bibinfo{author}{\bibfnamefont{J.~R.} \bibnamefont{{Shaw}}} \bibnamefont{and}
  \bibinfo{author}{\bibfnamefont{A.}~\bibnamefont{{Lewis}}},
  \bibinfo{journal}{\prd} \textbf{\bibinfo{volume}{86}}, \bibinfo{eid}{043510}
  (\bibinfo{year}{2012}).

\bibitem[{\citenamefont{{Yamazaki} et~al.}(2013)\citenamefont{{Yamazaki},
  {Ichiki}, and {Takahashi}}}]{2013PhRvD..88j3011Y}
\bibinfo{author}{\bibfnamefont{D.~G.} \bibnamefont{{Yamazaki}}},
  \bibinfo{author}{\bibfnamefont{K.}~\bibnamefont{{Ichiki}}}, \bibnamefont{and}
  \bibinfo{author}{\bibfnamefont{K.}~\bibnamefont{{Takahashi}}},
  \bibinfo{journal}{\prd} \textbf{\bibinfo{volume}{88}}, \bibinfo{eid}{103011}
  (\bibinfo{year}{2013}).

\bibitem[{\citenamefont{{Paoletti} and {Finelli}}(2013)}]{2013PhLB..726...45P}
\bibinfo{author}{\bibfnamefont{D.}~\bibnamefont{{Paoletti}}} \bibnamefont{and}
  \bibinfo{author}{\bibfnamefont{F.}~\bibnamefont{{Finelli}}},
  \bibinfo{journal}{\plb} \textbf{\bibinfo{volume}{726}}, \bibinfo{pages}{45}
  (\bibinfo{year}{2013}).

\bibitem[{\citenamefont{{Shiraishi}}(2013)}]{2013JCAP...11..006S}
\bibinfo{author}{\bibfnamefont{M.}~\bibnamefont{{Shiraishi}}},
  \bibinfo{journal}{\jcap} 
  \bibinfo{volume}{11} (\bibinfo{year}{2013}) \bibinfo{eid}{006}.

\bibitem[{\citenamefont{{Planck Collaboration}
  et~al.}(2016{\natexlab{a}})\citenamefont{{Planck Collaboration}, {Ade},
  {Aghanim}, {Arnaud}, {Arroja}, {Ashdown}, {Aumont}, {Baccigalupi},
  {Ballardini}, {Banday} et~al.}}]{2016A&A...594A..19P}
  \bibinfo{author}{\bibfnamefont{P.~A.~R.} \bibnamefont{{Ade}}},
  \bibinfo{author}{\bibfnamefont{N.}~\bibnamefont{{Aghanim}}},
  \bibinfo{author}{\bibfnamefont{M.}~\bibnamefont{{Arnaud}}},
  \bibinfo{author}{\bibfnamefont{F.}~\bibnamefont{{Arroja}}},
  \bibinfo{author}{\bibfnamefont{M.}~\bibnamefont{{Ashdown}}},
  \bibinfo{author}{\bibfnamefont{J.}~\bibnamefont{{Aumont}}},
  \bibinfo{author}{\bibfnamefont{C.}~\bibnamefont{{Baccigalupi}}},
  \bibinfo{author}{\bibfnamefont{M.}~\bibnamefont{{Ballardini}}},
  \bibinfo{author}{\bibfnamefont{A.~J.} \bibnamefont{{Banday}}},
  \bibnamefont{$et~al.$} (\bibinfo{author}{\bibnamefont{{Planck Collaboration}}}),
  \bibinfo{journal}{\aap} \textbf{\bibinfo{volume}{594}},
  \bibinfo{eid}{A19} (\bibinfo{year}{2016}{\natexlab{a}}).

\bibitem[{\citenamefont{{Zucca} et~al.}(2017)\citenamefont{{Zucca}, {Li}, and
  {Pogosian}}}]{2017PhRvD..95f3506Z}
\bibinfo{author}{\bibfnamefont{A.}~\bibnamefont{{Zucca}}},
  \bibinfo{author}{\bibfnamefont{Y.}~\bibnamefont{{Li}}}, \bibnamefont{and}
  \bibinfo{author}{\bibfnamefont{L.}~\bibnamefont{{Pogosian}}},
  \bibinfo{journal}{\prd} \textbf{\bibinfo{volume}{95}}, \bibinfo{eid}{063506}
  (\bibinfo{year}{2017}).

\bibitem[{\citenamefont{{Yamazaki}}(2014)}]{2014PhRvD..89j3528Y}
\bibinfo{author}{\bibfnamefont{D.~G.} \bibnamefont{{Yamazaki}}},
  \bibinfo{journal}{\prd} \textbf{\bibinfo{volume}{89}}, \bibinfo{eid}{083528}
  (\bibinfo{year}{2014}).

\bibitem[{\citenamefont{{Yamazaki}}(2016)}]{2016PhRvD..93d3004Y}
\bibinfo{author}{\bibfnamefont{D.~G.} \bibnamefont{{Yamazaki}}},
  \bibinfo{journal}{\prd} \textbf{\bibinfo{volume}{93}}, \bibinfo{eid}{043004}
  (\bibinfo{year}{2016}).

\bibitem[{\citenamefont{{Planck Collaboration}
  et~al.}(2016{\natexlab{b}})\citenamefont{{Planck Collaboration}, {Ade},
  {Aghanim}, {Arnaud}, {Ashdown}, {Aumont}, {Baccigalupi}, {Banday},
  {Barreiro}, {Bartlett} et~al.}}]{2016A&A...594A..13P}
\bibinfo{author}{\bibnamefont{{Planck Collaboration}}},
  \bibinfo{author}{\bibfnamefont{P.~A.~R.} \bibnamefont{{Ade}}},
  \bibinfo{author}{\bibfnamefont{N.}~\bibnamefont{{Aghanim}}},
  \bibinfo{author}{\bibfnamefont{M.}~\bibnamefont{{Arnaud}}},
  \bibinfo{author}{\bibfnamefont{M.}~\bibnamefont{{Ashdown}}},
  \bibinfo{author}{\bibfnamefont{J.}~\bibnamefont{{Aumont}}},
  \bibinfo{author}{\bibfnamefont{C.}~\bibnamefont{{Baccigalupi}}},
  \bibinfo{author}{\bibfnamefont{A.~J.} \bibnamefont{{Banday}}},
  \bibinfo{author}{\bibfnamefont{R.~B.} \bibnamefont{{Barreiro}}},
  \bibinfo{author}{\bibfnamefont{J.~G.} \bibnamefont{{Bartlett}}},
  \bibnamefont{$et~al.$}, \bibinfo{journal}{\aap} \textbf{\bibinfo{volume}{594}},
  \bibinfo{eid}{A13} (\bibinfo{year}{2016}{\natexlab{b}}).

\bibitem[{\citenamefont{Lewis et~al.}(2000)\citenamefont{Lewis, Challinor, and
  Lasenby}}]{camb}
\bibinfo{author}{\bibfnamefont{A.}~\bibnamefont{Lewis}},
  \bibinfo{author}{\bibfnamefont{A.}~\bibnamefont{Challinor}},
  \bibnamefont{and} \bibinfo{author}{\bibfnamefont{A.}~\bibnamefont{Lasenby}},
  \bibinfo{journal}{Astrophys. J.} \textbf{\bibinfo{volume}{538}},
  \bibinfo{pages}{473} (\bibinfo{year}{2000}).

\bibitem[{\citenamefont{Jedamzik et~al.}(1998)\citenamefont{Jedamzik,
  Katalinic, and Olinto}}]{Jedamzik:1996wp}
\bibinfo{author}{\bibfnamefont{K.}~\bibnamefont{Jedamzik}},
  \bibinfo{author}{\bibfnamefont{V.}~\bibnamefont{Katalinic}},
  \bibnamefont{and} \bibinfo{author}{\bibfnamefont{A.~V.}
  \bibnamefont{Olinto}}, \bibinfo{journal}{Phys. Rev.}
  \textbf{\bibinfo{volume}{D 57}}, \bibinfo{pages}{3264}
  (\bibinfo{year}{1998}).

\bibitem[{\citenamefont{Subramanian and
  Barrow}(1998{\natexlab{a}})}]{Subramanian:1997gi}
\bibinfo{author}{\bibfnamefont{K.}~\bibnamefont{Subramanian}} \bibnamefont{and}
  \bibinfo{author}{\bibfnamefont{J.~D.} \bibnamefont{Barrow}},
  \bibinfo{journal}{Phys. Rev.} \textbf{\bibinfo{volume}{D 58}},
  \bibinfo{pages}{083502} (\bibinfo{year}{1998}{\natexlab{a}}).

\bibitem[{\citenamefont{Subramanian and
  Barrow}(1998{\natexlab{b}})}]{Subramanian:1998fn}
\bibinfo{author}{\bibfnamefont{K.}~\bibnamefont{Subramanian}} \bibnamefont{and}
  \bibinfo{author}{\bibfnamefont{J.~D.} \bibnamefont{Barrow}},
  \bibinfo{journal}{Phys. Rev. Lett.} \textbf{\bibinfo{volume}{81}},
  \bibinfo{pages}{3575} (\bibinfo{year}{1998}{\natexlab{b}}).

\bibitem[{\citenamefont{{Yamazaki} and {Kusakabe}}(2012)}]{2012PhRvD..86l3006Y}
\bibinfo{author}{\bibfnamefont{D.~G.} \bibnamefont{{Yamazaki}}}
  \bibnamefont{and}
  \bibinfo{author}{\bibfnamefont{M.}~\bibnamefont{{Kusakabe}}},
  \bibinfo{journal}{\prd} \textbf{\bibinfo{volume}{86}}, \bibinfo{eid}{123006}
  (\bibinfo{year}{2012}).

\bibitem[{\citenamefont{Peebles}(1980)}]{Peebles:1980booka}
\bibinfo{author}{\bibfnamefont{P.~J.~E.} \bibnamefont{Peebles}},
  \emph{\bibinfo{title}{The Large-Scale Structure of the Universe}}
  (\bibinfo{publisher}{Princeton University Press, Princeton, NJ},
  \bibinfo{year}{1980}).

\bibitem[{\citenamefont{Kim et~al.}(1996)\citenamefont{Kim, Olinto, and
  Rosner}}]{Kim:1994zh}
\bibinfo{author}{\bibfnamefont{E.-J.} \bibnamefont{Kim}},
  \bibinfo{author}{\bibfnamefont{A.}~\bibnamefont{Olinto}}, \bibnamefont{and}
  \bibinfo{author}{\bibfnamefont{R.}~\bibnamefont{Rosner}},
  \bibinfo{journal}{Astrophys. J.} \textbf{\bibinfo{volume}{468}},
  \bibinfo{pages}{28} (\bibinfo{year}{1996}).

\bibitem[{\citenamefont{Tashiro and Sugiyama}(2006)}]{Tashiro:2005ua}
\bibinfo{author}{\bibfnamefont{H.}~\bibnamefont{Tashiro}} \bibnamefont{and}
  \bibinfo{author}{\bibfnamefont{N.}~\bibnamefont{Sugiyama}},
  \bibinfo{journal}{\mnras} \textbf{\bibinfo{volume}{368}},
  \bibinfo{pages}{965} (\bibinfo{year}{2006}).

\bibitem[{\citenamefont{{Meszaros}}(1974)}]{1974A&A....37..225M}
\bibinfo{author}{\bibfnamefont{P.}~\bibnamefont{{Meszaros}}},
  \bibinfo{journal}{\aap} \textbf{\bibinfo{volume}{37}}, \bibinfo{pages}{225}
  (\bibinfo{year}{1974}).

\bibitem[{\citenamefont{Ma and Bertschinger}(1995)}]{Ma:1995ey}
\bibinfo{author}{\bibfnamefont{C.-P.} \bibnamefont{Ma}} \bibnamefont{and}
  \bibinfo{author}{\bibfnamefont{E.}~\bibnamefont{Bertschinger}},
  \bibinfo{journal}{Astrophys. J.} \textbf{\bibinfo{volume}{455}},
  \bibinfo{pages}{7} (\bibinfo{year}{1995}).

\bibitem[{\citenamefont{{Hu} and {Sugiyama}}(1996)}]{1996ApJ...471..542H}
\bibinfo{author}{\bibfnamefont{W.}~\bibnamefont{{Hu}}} \bibnamefont{and}
  \bibinfo{author}{\bibfnamefont{N.}~\bibnamefont{{Sugiyama}}},
  \bibinfo{journal}{\apj} \textbf{\bibinfo{volume}{471}}, \bibinfo{pages}{542}
  (\bibinfo{year}{1996}).

\bibitem[{\citenamefont{{Yamamoto} et~al.}(1998)\citenamefont{{Yamamoto},
  {Sugiyama}, and {Sato}}}]{Yamamoto:1997qc}
\bibinfo{author}{\bibfnamefont{K.}~\bibnamefont{{Yamamoto}}},
  \bibinfo{author}{\bibfnamefont{N.}~\bibnamefont{{Sugiyama}}},
  \bibnamefont{and} \bibinfo{author}{\bibfnamefont{H.}~\bibnamefont{{Sato}}},
  \bibinfo{journal}{\apj} \textbf{\bibinfo{volume}{501}}, \bibinfo{pages}{442}
  (\bibinfo{year}{1998}).

\bibitem[{\citenamefont{{Challinor} and {Lewis}}(2005)}]{2005PhRvD..71j3010C}
\bibinfo{author}{\bibfnamefont{A.}~\bibnamefont{{Challinor}}} \bibnamefont{and}
  \bibinfo{author}{\bibfnamefont{A.}~\bibnamefont{{Lewis}}},
  \bibinfo{journal}{\prd} \textbf{\bibinfo{volume}{71}}, \bibinfo{eid}{103010}
  (\bibinfo{year}{2005}).

\bibitem[{\citenamefont{{Lewis} and {Challinor}}(2006)}]{2006PhR...429....1L}
\bibinfo{author}{\bibfnamefont{A.}~\bibnamefont{{Lewis}}} \bibnamefont{and}
  \bibinfo{author}{\bibfnamefont{A.}~\bibnamefont{{Challinor}}},
  \bibinfo{journal}{\physrep} \textbf{\bibinfo{volume}{429}},
  \bibinfo{pages}{1} (\bibinfo{year}{2006}).

\bibitem[{\citenamefont{{Silk}}(1968)}]{1968ApJ...151..459S}
\bibinfo{author}{\bibfnamefont{J.}~\bibnamefont{{Silk}}},
  \bibinfo{journal}{Astrophys. J.} \textbf{\bibinfo{volume}{151}},
  \bibinfo{pages}{459} (\bibinfo{year}{1968}).

\end{thebibliography}

\end{document}